\documentclass{article}
\usepackage{fleqn,espcrc2}
\usepackage{graphicx}
\usepackage[figuresright]{rotating}

\newcommand{\AmS}
{{\protect\the\textfont2A\kern-.1667em\lower.5ex\hbox{M}\kern-.125emS}}
\title{Inelastic scattering rates in d-wave superconductors.}

\author{ J. Paaske
\address{{\O}rsted Laboratory,
         University of Copenhagen,
         Universitetsparken 5, DK-2100 Copenhagen, Denmark.}
         and 
         D.V. Khveshchenko
\address{Department of Physics and Astronomy, University of North Carolina,
         Chapel Hill, NC 27599, USA.}}

\begin{document}

\begin{abstract}
The inelastic scattering rates of quasiparticles in a two-dimensional d-wave
superconductor, which arise from interactions with either acoustic phonons or
other quasiparticles, are calculated within second order perturbation theory.
We discover a strong enhancement of scattering with collinear momenta, brought
about by the special kinematics of the two-dimensional fermions with Dirac-like
spectrum near the nodes of the d-wave order parameter. In the case of a local
instantaneous interparticle potential we find that either an RPA-type resummation
of the perturbation series or an inclusion of non-linear corrections to the Dirac
spectrum is called for in order to obtain a finite scattering rate in the limit
$\omega/T\rightarrow 0$. In either way, we find drastic changes in the scattering
rate, as compared to the naively expected cubic temperature dependence.
\vspace{1pc}
\end{abstract}

\maketitle


Recent measurements of microwave~\cite{Hosseini99} and thermal
conductivity~\cite{Krishana99} in 90 K YBCO have probed the quasiparticle
relaxation rates in a range of temperatures below $T_{c}$ for which inelastic
scattering is believed to be important, and rates approximately proportional
to $T^{4}$ and $T^{2}$, respectively, were revealed. Interestingly enough, both 
behaviours deviate from the simple cubic rate which would follow from the naive
Golden rule estimate for the el-el scattering rate in the presence of a linear
density of quasiparticle states. The likelihood of this discrepancy being due to
an intricate difference between the  quasiparticle lifetime versus, generally
different, charge and energy relaxation rates motivates one to revisit the
calculations based upon the golden rule.

We first consider the scattering of quasiparticles off three-dimensional acoustic
phonons at temperatures below the Debye temperature. After having integrated the
second order perturbation theory result for the inelastic quasiparticle lifetime
over the out-of-plane phonon momenta, one recovers the expected cubic rate
\begin{equation}
\tau_{el-ph}^{-1}\sim I(1)\frac{T^{3}}{\theta_{Debye}\Delta_{0}},
\end{equation}
which is frequency independent in the limit $\omega\ll T$. However, the prefactor
$I(1)$ obtained from the angular part of the planar momentum integral becomes
divergent when the speed of sound $s$ approaches the value of the quasiparticle
velocity $v_{2}$ parallel to the Fermi surface (the latter is proportional to the
maximum superconducting gap and therefore much smaller than $v_{F}$):
$I(1)\sim \ln(v_{2}/|v_{2}-s|)$. 

Although posing no real problem for the el-ph scattering rate, the above divergence
reflects a potential danger in the case of el-el scattering of quasiparticles with
equal velocities.

To this end, we consider a local instantaneous interaction between either charge or
spin densities. In the leading approximation, one can neglect scattering processes 
between any pair of the neighbouring nodes because of the strong eccentricity of
the equal energy contours in momentum space
($v_{F}/v_{2}\sim 14\pm 3$)~\cite{Chiao99}. For the remaining {\it intra-node} and
{\it opposite-node} processes scaling of the local momenta near the nodes by the
corresponding velocities gives rise to the symmetrical Dirac-like quasiparticle
spectrum $E({\bf k})=|{\bf k}|$. 

Further analysis reveals, that amongst the terms corresponding to different
combinations of coherence factors~\cite{Schrieffer93} and nodes, involved in the
scattering processes, the most important ones are those which correspond to
scattering off thermally excited quasiparticles with nearly parallel momenta.

It proves convenient to write the scattering rate in terms of the charge or spin
susceptibility
\begin{eqnarray}
\chi^{\prime\prime}({\bf q},\Omega)
&=&\frac{1}{v_{1}v_{2}}\int\!\frac{d^{2}k'}{(2\pi)^{2}}
\left[f(k'-\Omega)-f(k')\right]
\nonumber\\
&&\hspace*{15.4mm}\times\,
\delta(\Omega-k'-|{\bf k}'+{\bf q}|)\nonumber\\
&\approx&\frac{\theta(\Omega-q)}{64\,v_{1}v_{2}}\,
\frac{2\Omega^{2}-q^{2}}{\sqrt{\Omega^{2}-q^{2}}}\,\frac{\Omega}{T},
\label{eq:suscept}
\end{eqnarray}
valid for $\Omega\ll T$. Eq. (\ref{eq:suscept}) is then integrated with the
energy conserving delta function and the appropriate combination of distribution
functions, and for the external momentum taken right on the node this yields the
integral 
\begin{eqnarray}
\tau^{-1}(\omega)&\!\!\sim&\!\!
\int_{0}^{\infty}\!\!\!dq\,
\frac{q^{2}(2(\omega+q)^{2}-q^{2})}{v_{1}v_{2}T\sqrt{\omega(\omega+2q)}}\,
{\rm csch}\left(\frac{q}{T}\right)\nonumber\\
&\!\!\sim&\!\!\frac{T^{7/2}}{t\Delta_{0}\omega^{1/2}},
\hspace*{5mm}{\rm for}\hspace*{3mm}\omega\ll T,\label{eq:rate}
\end{eqnarray}
which tends to diverge in the limit $\omega/T\rightarrow 0$, unlike the omitted
regular terms proportional to $\omega^{3}$ or $\omega T^{2}$. A mere substitution
of $\omega$ by $T$ would yield a cubic temperature dependence but conceal the
divergence stemming from the momenta $q\approx\Omega$. This kinematical singularity,
which reflects enhancement of scattering between particles with collinear momenta,
has also been noted in the early studies of semi-metals~\cite{Abrikosov71}.

In lattice models, the above divergence is readily cut off by non-linear corrections
to the bare quasiparticle spectrum. In the case of parameters relevant for the
problem of high temperature superconductors ($t'\sim -0.5 t$) we obtain 
\begin{equation}
\tau^{-1}\sim\left(\frac{T}{t}\right)^{3/2}T.\label{eq:nonlin}
\end{equation}

Nonetheless, in the case of a strictly linear spectrum no physically meaningful
result can be obtained without a summation of the entire perturbation series.
As an attempt to carry out this procedure one can substitute the bare interaction
$\lambda$ by an effective interaction
\begin{equation}
D(i\nu,{\bf q})=\frac{\lambda}{1-\lambda\chi(i\nu,{\bf q})},
\end{equation}
containing the singular susceptibility (\ref{eq:suscept}). In the limit
$\omega/T\ll (\lambda T/t\Delta_{0})^{2}$, this procedure yields the rate
\begin{equation}
\tau^{-1}\sim
\left[\left(\frac{\lambda T}{t\Delta_{0}}\right)^{3}\frac{\omega}{T}\right]^{1/5}T,
\end{equation}
and in general there might be other regimes of higher $\omega$ leading to
different temperature dependences. Observe that neither this rate nor
(\ref{eq:nonlin}) reverts to a cubic temperature dependence when substituting
$T$ for $\omega$. 

An important exclusion from the above generic situation is provided by the case of
el-el interactions mediated by antiferromagnetic spin fluctuations, whose ordering
vector is commensurate with the distance between the opposite pairs of nodes. In
this case one arrives at a non-singular second order result
$\tau^{-1}\sim\omega^{1/2}\,T^{5/2}$.

Although it is widely believed that this particular channel is relevant for the
cuprates, it remains to be seen whether the lowest order estimate captures all the
relevant physics. A formally related example of the system of interacting Dirac
fermions describing quantum Hall plateau transitions ~\cite{Sachdev97} indicates
that it may indeed be the case.

J.P. gratefully acknowledges P.A. Lee for many stimulating discussions of
this problem.



\begin{thebibliography}{9}

\bibitem{Hosseini99} A. Hosseini {\it et al.},
Phys. Rev. B 60 (1999) 1349.

\bibitem{Krishana99} K. Krishana {\it et al.},
Phys. Rev. Lett. 82 (1999) 5108.

\bibitem{Chiao99} M. Chiao {\it et al.},
Phys. Rev. Lett. 82 (1999) 2943.

\bibitem{Schrieffer93} J.R. Schrieffer,
{\em Theory of Superconductivity}, 4. printing, Addison-Wesley (1993).

\bibitem{Abrikosov71} A.A. Abrikosov, S.D. Beneslavskii,
Sov. Phys. JETP 32 (1971) 699.

\bibitem{Sachdev97} S. Sachdev,
Phys. Rev. B 57 (1998) 7157.

\end{thebibliography}
\end{document}